\documentclass[12pt]{article}
\usepackage{amsmath, amssymb, mathrsfs, mathtools, tensor, braket}
\usepackage{xcolor, graphicx, subfigure}

\numberwithin{equation}{section}

\usepackage{enumitem, float}

\usepackage{caption}

\evensidemargin \oddsidemargin

\newcommand{\Romtwo}{I\(\!\)I}

\begin{document}

\begin{titlepage}

\begin{center}
\vspace{5mm}
    
{\Large \bfseries
	Vacuum and Vortices
	\\
	in Inhomogeneous Abelian Higgs Model
}\\[17mm]
    
	Yoonbai Kim,
	~~SeungJun Jeon,
	~~O-Kab Kwon,
	~~Hanwool Song
\\[2mm]
{\itshape
	Department of Physics,
	Sungkyunkwan University,
	Suwon 16419,
	Korea
}
\\[-1mm]
{\itshape
	yoonbai@skku.edu,
	~sjjeon@skku.edu,
	~okab@skku.edu,
	~hanwoolsong0@gmail.com
}
\\[3mm]
	Chanju Kim
\\[2mm]
{\itshape
	Department of Physics,
	Ehwa Womans University,
	Seoul 03760,
	Korea
}
\\[-1mm]
{\itshape
	cjkim@ewha.ac.kr
}

\end{center}
\vspace{15mm}

\begin{abstract}

\noindent
The inhomogeneous abelian Higgs model with a magnetic impurity in the BPS limit is studied for both relativistic and nonrelativistic regimes.
Though the symmetry of spatial translation is broken by inhomogeneity, extension to an \(\mathcal{N}=1\) supersymmetric theory is admitted.
The quartic scalar potential has minimum value depending on strength of the impurity but possesses broken phase at spatial asymptote.
The vacuum configuration of broken phase can be neither a constant nor the minimum of the scalar potential, but is found as a nontrivial solution of the Bogomolny equations.
While its energy density and magnetic field are given by the function of spatial coordinates, the energy and magnetic flux remain zero.
The sign of the magnetic impurity term allows either a BPS sector or anti-BPS sector but not both.
Thus the obtained solution is identified as the new inhomogeneous broken vacuum of minimum zero energy.
In the presence of rotationally symmetric Gaussian type inhomogeneity, topological vortex solutions are also obtained and the effects of the impurity to the vortex are numerically analyzed.

\end{abstract}

\end{titlepage}

\section{Introduction}

Field theories have been regarded as a most suitable tool for describing the fundamental forces in microscopic level.
On the other hand, samples in real experiments involve diversified impurities, defects, disorders, etc., by doping, imperfect growth of samples, junctions of heterogeneous materials, etc.
Hence field theoretic description of these samples immediately encounters difficulty in analytic treatment and an adequate guideline is necessary.
Since majority of field theories consist of the fields and constant parameters such as masses and couplings at least at the textbook level, an available and simple window is to allow inhomogeneity or spatial dependence, in one or a few parameters.
Then the field theories with these additional ingredients become complicated as usual and another controllable guideline is indispensable for tractability.
A familiar option in field theories is supersymmetry.
Even if a part of spacetime symmetries, the Poincar{\'e} symmetry or the Galilean symmetry, is explicitly broken by the presence of inhomogeneity, a reduced number of supersymmetries are known to be survived.
These so-called inhomogeneous field theories have begun with the name of Janus in supersymmetric and non-supersymmetric field theories under the supervision of holography \cite{Bak:2003jk, Clark:2005te, DHoker:2006qeo, DHoker:2007zhm, Kim:2008dj, Kim:2009wv}.
Then, mass-deformed ABJM theory in three dimensions and super Yang-Mills theory in three and four dimensions can allow inhomogeneous mass deformations in relation to the irregular form-fields on the branes, preserving same amount of supersymmetries \cite{Kim:2018qle, Kim:2019kns, Arav:2020obl, Kim:2020jrs}.
Solitonic excitations, kinks, are explored in two-dimensional supersymmetric theories including impurities \cite{Adam:2019yst, Adam:2018pvd, Adam:2018tnv, Adam:2019djg, Manton:2019xiq, Adam:2019hef, Adam:2019xuc} and then general form of the superpotential with spatial dependence for a single scalar field is identified and corresponding general solutions of the Bogomolny equation is obtained \cite{Kwon:2021flc}.
It has also been reported that, at the classical level, there is a one-to-one correspondence between supersymmetric inhomogeneous field theories in \((1+1)\) dimensions and supersymmetric field theories on a specific curved background metric \cite{Ho:2022omx}.
In relation with electromagnetism, inhomogeneous supersymmetric abelian gauge theories including electric and magnetic impurities have been considered in three and four dimensions and the effect of impurities on point charges and vortices are studied including vortex dynamics \cite{Hook:2013yda, Tong:2013iqa, Kim:2023abp}.

In this paper, we consider the abelian Higgs model with spatial inhomogeneity in both relativistic and nonrelativistic regime.
In scalar potential, quartic scalar coupling is chosen by the critical value which corresponds to equal London penetration depth and correlation length in homogeneous limit and admits the BPS bound.
If the inhomogeneity is introduced through the quadratic scalar term and the magnetic impurity term and is assumed to approach zero value at spatial infinity, the first observation is an explicit breakdown of the spatial translational symmetry.
Nevertheless, extension to an \(\mathcal{N}=1\) supersymmetric theory is allowed \cite{Hook:2013yda, Tong:2013iqa} and the BPS structure is preserved for both relativistic and nonrelativistic models.
A notable feature is that, depending on the sign of the magnetic impurity term, either a BPS sector or an anti-BPS sector is allowed, but not both.

The second observation is the absence of the constant symmetry-broken vacuum which is the well-known Higgs vacuum in the homogeneous limit.
Instead we find an inhomogeneous vacuum solution by examining the Bogomolny equations.
Its energy density and magnetic field are nontrivial functions of spatial coordinates, but the energy and magnetic flux are held to be zero.
Since the possible minimum energy for any static BPS configuration is zero according to the BPS structure for the inhomogeneous abelian Higgs model in both relativistic and nonrelativistic regime, the obtained solution deserves to be the new unique inhomogeneous, electrically neutral and spinless vacuum of minimum zero energy.

The abelian Higgs model supports topological vortex solutions called the Abrikosov-Nielsen-Olesen vortices, which carry quantized magnetic flux.
In the presence of inhomogeneity, we study the existence and properties of the BPS vortices and figure out the effect of impurity:
The magnetic flux and energy of static BPS vortices are unaffected but the shape of the magnetic field and energy densities are changed in the region of nonvanishing inhomogeneity.
Throughout this paper, our analyses are mostly carried out assuming a rotationally symmetric Gaussian-type inhomogeneity centered at the origin, and we obtain numerically the rotationally symmetric solutions of broken vacuum and \(n\)-superimposed vortices.

The rest of the paper is organized as follows:
In subsection 2.1, we introduce the relativistic abelian Higgs model added by a magnetic impurity term and derive the BPS bound including the an \(\mathcal{N}=1\) supersymmetric extension.
In subsection 2.2, similar analysis is done for the nonrelativistic abelian Higgs model with some discussions on superconductivity.
Section 3 is devoted to finding a new inhomogeneous broken vacuum solution of minimum energy.
In section 4, we obtain inhomogeneous topological vortex solutions and study the effect of inhomogeneity.
We conclude in section 5 with discussions.

\section{BPS Limit of Inhomogeneous Abelian Higgs Model}

\subsection{Relativistic model}

In \((1+2)\) dimensions with spacetime signature \((-,+,+)\), the abelian Higgs model is described by the Lagrangian density
\begin{equation}
	\mathscr{L}_\text{AH}
	=
	- \frac{1}{4} F_{\mu\nu} F^{\mu\nu}
	- \overline{D_{\mu}\phi} D^{\mu}\phi
	- V(|\phi|)
,\label{201}
\end{equation}
where \(\phi\) is a complex scalar field with covariant derivative
\(
	D_{\mu} \phi
	= (\partial_{\mu} - igA_{\mu})\phi
\).
If the scalar potential \(V\) is given by
\begin{equation}
	V(|\phi|)
	= \frac{g^2}{2}
	( |\phi|^{2} - v^{2} )^{2}
\label{202}
\end{equation}
whose quartic coupling is fixed by the gauge coupling \(g\), the BPS bound is saturated \cite{Bogomolny:1975de} and the model admits a \(\mathcal{N}=2\) supersymmetric extension \cite{DiVecchia:1977nxl, Witten:1978mh}.
With the non-zero vacuum expectation value \(\langle |\phi| \rangle=v\), the U(1) gauge symmetry of the theory is broken.
Then the gauge boson with both transverse and longitudinal degrees and a neutral Higgs boson propagate with degenerate mass
\begin{equation}
	m_{A_\mu}
	= \sqrt{2} gv
	= m_{\text{H}}
.\label{203}
\end{equation}

In this paper we impose spatial coordinates dependence on the parameter \(v\) and assume that \(v\) approaches a constant vacuum expectation value \(v_0\) at spatial infinity
\begin{equation}
	v^{2}(\boldsymbol{x})
	\longrightarrow
	v_{0}^{2}
,\qquad
	r \equiv |\boldsymbol{x}|
	\longrightarrow
	\infty
, \qquad
	\boldsymbol{x}
	= (x_{1} ,  x_{2})
.\label{224}
\end{equation}
Let us define \( \sigma(\boldsymbol{x}) \) called inhomogeneity by
\begin{equation}
	v^{2} (\boldsymbol{x} )
	= v_{0}^{2} + \sigma(\boldsymbol{x})
,\qquad
	\lim_{|\boldsymbol{x}|\to\infty}
	\sigma(\boldsymbol{x})
	= 0
.\label{204}
\end{equation}
Such inhomogeneity is probably originated from impurities in the system of consideration or from heterogeneous systems being joined together.
If \( \sigma(\boldsymbol{x}) \) is a nontrivial function, a part of Poincar{\'e} symmetry is explicitly broken.
As discussed below, however, the BPS nature can be kept if another inhomogeneous term called the magnetic impurity term is added to the Lagrangian,
\begin{equation}
	\Delta \mathscr{L}
	= s g \sigma (\boldsymbol{x}) B
,\label{205}
\end{equation}
where \(s\) is either \(+1\) or \(-1\) and \(B\) is the magnetic field
\(
	B =
	\epsilon_{ij} \partial_{i} A_{j}
\).
This magnetic impurity term has been introduced in the context of abelian Higgs model as a supersymmetry-preserving impurity \cite{Tong:2013iqa, Hook:2013yda}.
Origin of the magnetic impurity can be reconciled through field theories of larger gauge group, e.g.
\(
	\mathrm{U}(1)
	\times
	\mathrm{U}(1)
\)
gauge model with two complex scalar fields, in which the magnetic impurities are realized as vortices in the infinitely heavy limit after integrating out the heavy scalar field with the help of BPS equations \cite{Tong:2013iqa}.
Now the Lagrangian density of our consideration is
\begin{equation}
	\mathscr{L}
	=
	- \frac{1}{4} F_{\mu\nu} F^{\mu\nu}
	- \overline{D_{\mu}\phi} D^{\mu}\phi
	- \frac{g^2}{2} (
		|\phi|^{2} - v^{2} (\boldsymbol{x})
	)^{2}
	+ sg \sigma(\boldsymbol{x}) B
.\label{206}
\end{equation}

We read the energy from the Lagrangian density
\begin{equation}
	E =
	\int d^{2} x \,
	\bigg[
		\frac{1}{2} \boldsymbol{E}^{2}
		+ \frac{1}{2} B^{2}
		+ |D_{0} \phi|^{2}
		+ |D_{i} \phi|^{2}
		+ \frac{g^{2}}{2}
		( |\phi|^{2} - v^{2} (\boldsymbol{x}) )^{2}
		- sg \sigma B
	\bigg]
,\label{221}
\end{equation}
and reshuffle the terms by application of the Bogomolny trick for static configurations as
\begin{equation}
	E =
	\int d^{2} x \,
	\bigg[
		|(D_{1} + is D_{2}) \phi|^{2}
		+ \frac{1}{2}
		| B + sg (|\phi|^{2} - v^{2} (\boldsymbol{x}) ) |^{2}
		+ sg v_{0}^{2} B
	\bigg]
,\label{207}
\end{equation}
where a vanishing surface term is neglected and the last term is obtained by cancellation of the remaining inhomogeneous parts in the last two terms in \eqref{221}.
In the middle of derivation of this expression, the relation
\(
	[ D_{1}, D_{2} ]
	= -ig F_{12}
	= -ig B
\)
has been used and Weyl gauge condition \( A^{0} = 0 \) is chosen.
Then any static configuration does not carry non-zero electric field
\begin{equation}
	E_{i} = F_{i0} = 0
.\label{226}
\end{equation}

Notice that the energy is bounded from below by the magnetic flux
\begin{equation}
	E
	\ge
	sg v_{0}^{2} \int d^{2} x \, B
	=
	sg v_{0}^{2} \Phi_{B}
,\label{208}
\end{equation}
and the bound is saturated if the following Bogomolny equations hold,
\begin{align}
	(D_{1} + is D_{2} ) \phi
	& = 0
,\label{209}
\\
	B + sg ( |\phi|^{2} - v^{2}(\boldsymbol{x}) )
	& = 0
.\label{210}
\end{align}
It is straightforward to check that every static solution of these equations automatically satisfies the second-order Euler-Lagrange equations.
Once the sign \(s\) of the inhomogeneous term \eqref{205} is fixed, so are the energy bound \eqref{208} as well as the Bogomolny equations.
In the usual homogeneous limit of \(\sigma(\boldsymbol{x})=0\), both \(s=1\) and \(s=-1\) are allowed when completing the squares in the single model, leading two separate energy bounds accordingly and hence we have nonnegative energy
\(
	E \ge
	v_{0}^{2} |g\Phi_{B}|
\).
In the present case, however, only one energy bound holds because it is tied to the sign in front of the inhomogeneous term.
We will fix \(s=1\) from now on without loss of generality since one can obtain \(s=-1\) case under the parity transformation, \(x_{2} \to -x_{2}\) and \(A_{2} \to -A_{2}\).

For any static configuration, the momentum density \(T_{0i}\) vanishes 
\begin{equation}
	T_{0i}
	= 
	F_{0j} F\indices{_i^j}
	+ \overline{D_{0} \phi} D_{i} \phi
	+ \overline{D_{i} \phi} D_{0} \phi
	= 0
.\label{228}
\end{equation}
Thus every static BPS configuration is left to be spinless 
\(	\displaystyle
	J =
	\int d^{2}x \, \epsilon^{ij} x_{i} T_{0j}
	= 0
\)
despite of the inhomogeneity \( \sigma(\boldsymbol{x}) \).
The stress components \(T_{ij}\) of symmetrized energy-momentum tensor can be written through reshuffling as
\begin{equation}
\begin{aligned}[b]
	T_{ij}
	& =
	\frac{1}{2}
	[ B - g ( |\phi|^{2} - v^{2} ) ]
	[ B + g ( |\phi|^{2} - v^{2} ) ]
	\delta_{ij}
\\ 
	& \hphantom{=}
	+ \frac{1}{4}
	\Big[
		\overline{
			(D_{i} - i\epsilon^{ik} D_{k}) \phi
		} (D_{j} + i\epsilon^{jl} D_{l}) \phi
		+
		\overline{
			(D_{i} + i\epsilon^{ik} D_{k}) \phi
		} (D_{j} - i\epsilon^{jl} D_{l}) \phi
\\
	& \qquad \quad
	+
		\overline{
			(D_{j} - i\epsilon^{jk} D_{k}) \phi
		} (D_{i} + i\epsilon^{il} D_{l}) \phi
		+
		\overline{
			(D_{j} + i\epsilon^{jk} D_{k}) \phi
		} (D_{i} - i\epsilon^{il} D_{l}) \phi
	\Big]
,\label{211}
\end{aligned}
\end{equation}
which vanishes upon applying the Bogomolny equations.
In addition, it readily leads to modification of the momentum conservation
\begin{equation}
	\partial_{\mu} T^{\mu \nu}
	=
	g [ B + g ( |\phi|^{2} - v^{2} )]
	\partial^{\nu} \sigma (\boldsymbol{x})
.\label{212}
\end{equation}
Therefore, the momentum is not conserved. Of course, this is because the inhomogeneity breaks the translation symmetry.
Note that the right hand side is proportional to the Bogomolny equation \eqref{210} which vanishes for BPS solutions, so that it is consistent with \eqref{211}.

Though the Poincar{\'e} symmetry is explicitly broken, BPS nature of the model suggests that it still has a supersymmetric extension \cite{Tong:2013iqa, Adam:2019yst,Kwon:2021flc, Kim:2023abp}. 
In fact, it is precisely given by the \(\mathcal{N}=2\) supersymmetric homogeneous abelian Higgs model \cite{DiVecchia:1977nxl, Witten:1978mh} modified by the magnetic impurity term in \eqref{205} without further correction,
\begin{equation}
\begin{aligned}[b]
	\mathscr{L}_{\text{SUSY}}
	= &
	- \frac{1}{4} F_{\mu\nu} F^{\mu\nu}
	- \overline{D_{\mu} \phi} D^{\mu} \phi
	+ i\bar{\psi} \gamma^{\mu} D_{\mu} \psi
	- \frac{1}{2} \partial_{\mu} N \partial^{\mu} N
	+ i\bar{\chi} \gamma^{\mu} \partial_{\mu} \chi
\\	&
	+ i\sqrt{2} g (\bar{\psi} \chi \phi
	+ \bar{\chi} \psi \bar{\phi})
	+ ig \bar{\psi} \psi N
	- g^{2} N^{2} |\phi|^{2}
	- \frac{1}{2} g^{2}
	( |\phi|^{2} - v^{2} )^{2}
	+ g \sigma(\boldsymbol{x}) B
.
\end{aligned}
\label{213}
\end{equation}
Even in the presence of magnetic impurity term, the Lagrangian \eqref{213} is invariant up to a total derivative under the following supersymmetric transformation,
\begingroup \allowdisplaybreaks
\begin{align}
    \delta \phi
    & =
    i\sqrt{2} \bar{\varepsilon} \psi
, \nonumber \\
    \delta \bar{\phi}
    & =
    i\sqrt{2} \bar{\psi} \varepsilon
, \nonumber \\
    \delta N
    & =
    i (\bar{\varepsilon} \chi
    + \bar{\chi} \varepsilon)
, \nonumber \\
    \delta \psi
    & =
    - \sqrt{2} \gamma^{\mu} \varepsilon D_{\mu} \phi
    - \sqrt{2} g N \phi \varepsilon
, \nonumber \\
    \delta \bar{\psi}
    & =
    + \sqrt{2} \overline{D_{\mu} \phi} \bar{\varepsilon} \gamma^{\mu}
    - \sqrt{2} g N \bar{\phi} \bar{\varepsilon}
, \nonumber \\
    \delta \chi
    & =
    - \gamma^{\mu} \varepsilon \partial_{\mu} N
    + \frac{i}{2} \epsilon^{\mu\nu\rho} F_{\mu\nu}
    \gamma_{\rho} \varepsilon
    - g
        (
            |\phi|^{2} -v^{2}
        ) \varepsilon
, \nonumber \\
    \delta \bar{\chi}
    & =
    + \bar{\varepsilon} \gamma^{\mu} \partial_{\mu} N
    + \frac{i}{2} \epsilon^{\mu\nu\rho} F_{\mu\nu}
    \bar{\varepsilon} \gamma_{\rho}
    - g
        (
            |\phi|^{2} -v^{2}
        ) \bar{\varepsilon}
, \nonumber \\
    \delta A_{\mu}
    & =
    \bar{\varepsilon} \gamma_{\mu} \chi
    + \bar{\chi} \gamma_{\mu} \varepsilon
,\label{214}
\end{align}
\endgroup
provided that the complex parameter \(\varepsilon\) satisfies the condition
\begin{equation}
	\gamma^{1} \varepsilon
	= -i \gamma^{2} \varepsilon
.\label{215}
\end{equation}
Thus we have a reduced supersymmetry from \(\mathcal{N} = 2\) to \(\mathcal{N} = 1\) in inhomogeneous case.

With the projection condition \eqref{215}, the supersymmetric variation \(\delta\psi\), \(\delta\chi\) of the fermion fields in \eqref{214} can be written as,
\begin{align}
	\delta\psi
	& =
	- \sqrt{2} \gamma^{1} \varepsilon
	(D_{1} + iD_{2}) \phi
	+ \sqrt{2} i \varepsilon
	( D_{0} + igN ) \phi
, \\
	\delta\chi
	& =
	\varepsilon \big[
		- i \partial_{0} N
		+ B
		+ g (|\phi|^{2} - v^{2})
	\big]
	- \gamma^{i}\varepsilon
	(\partial_{i} N - E_{i})
,
\end{align}
which vanish if the following equations are satisfied
\begin{align}
	(D_{0} + i g N) \phi
	& = 0
,\label{219}
\\
	(D_{1} + i D_{2} ) \phi
	& = 0
,
\\
	- i\partial_{0} N 
	+ B + g
	    (       |\phi|^{2} -v^{2} 
	            (\boldsymbol{x})        )
	& = 0
,\label{222}
\\
	\partial_{i} N - E_{i}
	& = 0
.\label{220}
\end{align}
For any static BPS configuration, time derivative terms vanish in \eqref{219}, \eqref{222}, and \eqref{220}.
Thus the equations \eqref{219} and \eqref{220} reduce to a relation \(A^{0} = N\) and the remaining two first-order equations become identical to the Bogomolny equations \eqref{209} and \eqref{210}, as it should be.
The unbroken supercharges of the model can be obtained by a standard procedure,
\begingroup
\allowdisplaybreaks
\begin{align}
    Q = &\, \int d^{2} x \,
    \Big\{
            - i \big[
                (\partial_{1} N - E_{1})
                - i (\partial_{2} N - E_{2})
            \big] \chi_{1}^{\dagger}
            + \big[
                - i \partial_{0} N
                + B
                + g (|\phi|^{2} - v^{2})
            \big] \chi_{1}^{\dagger}
\nonumber \\
& \hspace{4em}
            + \big[
                (\partial_{1} N - E_{1})
                - i (\partial_{2} N - E_{2})
            \big] \chi_{2}^{\dagger}
            + i \big[
                -i \partial_{0} N
                + B
                + g (|\phi|^{2} - v^{2})
                \big] \chi_{2}^{\dagger}
\nonumber \\
& \hspace{4em}
            - i \big[
                \sqrt{2} (D_{0} + i g N) \phi
                + \sqrt{2} (D_{1} + i D_{2}) \phi
            \big] \psi_{1}^{\dagger}
\nonumber\\
& \hspace{4em}
            + \big[
                \sqrt{2} (D_{0} + ig N) \phi
                - \sqrt{2} (D_{1} + i D_{2}) \phi
            \big] \psi_{2}^{\dagger}
    \Big\}
,\nonumber \\
    Q^{\dagger} =&\, \int d^{2} x\,
    \Big\{
        i \big[
                (\partial_{1} N - E_{1})
                + i (\partial_{2} N - E_{2})
            \big] \chi_{1}
            + \big[
                i \partial_{0} N
                + B
                + g (|\phi|^{2} - v^{2})
            \big] \chi_{1}
\nonumber\\
& \hspace{4em}
        + \big[
                (\partial_{1} N - E_{1})
                + i (\partial_{2} N - E_{2})
        \big] \chi_{2}
        - i \big[
                i \partial_{0} N
                + B
                + g (|\phi|^{2} - v^{2})
        \big] \chi_{2} 
\nonumber\\
& \hspace{4em}
                + i \big[
                        \sqrt{2}\, \overline{(D_{0} + i g N) \phi}
                        + \sqrt{2}\, \overline{(D_{1} + i D_{2}) \phi}
                \big] \psi_{1}
\nonumber\\
& \hspace{4em}
                + \big[
                        \sqrt{2}\, \overline{(D_{0} + ig N) \phi}
                        - \sqrt{2}\, \overline{(D_{1} + i D_{2}) \phi}
                \big] \psi_{2}
        \Big\}
,\label{216} 
\end{align}
\endgroup
and satisfy the superalgebra
\begin{equation}
\begin{aligned}[b]
        \{ Q, Q \} 
        & =
        \{ Q^{\dagger}, Q^{\dagger} \}
        = 0
, \\
        \{ Q, Q^{\dagger} \}
        & =
        4E + 4 \int d^{2} x \,
    \bigg[
        E^{i} \partial_{i} N
        - i g N (
                \bar{\phi} D_{0} \phi
                - \overline{D_{0} \phi} \phi )
        - \partial_{i} \Big(
                -\frac{1}{2g} \epsilon^{ij} j_{j}
        \Big)
        - g v_{0}^{2} B
    \bigg]
\\
        & =
        4 ( E - gv_{0}^{2} \Phi_{B} )
.\label{217}
\end{aligned}
\end{equation}
The last result \eqref{217} reproduces the energy bound \eqref{208}.

Let us focus on the Bogomolny equations \eqref{209}--\eqref{210}.
It is well-known that if \(\phi\) is nonvanishing and satisfies \eqref{209} with \(s=1\), then its zeros, if it has any, are isolated and only positive vortex number is allowed, i.e., \( \Phi_{B} \ge 0 \) \cite{Jaffe:1980mj}.
Therefore, the model can have BPS solutions only with positive vorticities, in contrast to the homogeneous case in which both BPS vortices and BPS antivortices exist.
Let \(\boldsymbol{x}_{a}\) with \(a = 1, 2, \dots, n\) be the zeros of complex scalar field \(\phi\).
Eliminating the gauge field, it is straightforward to combine the two equations \eqref{209}--\eqref{210} into a single second-order equation
\begin{equation}
    \nabla^{2} \ln |\phi|^{2}
    =
    2g^{2} [ |\phi|^{2} - v^{2} (\boldsymbol{x}) ]
    +
    4 \pi \sum_{a=1}^{n}
    \delta( \boldsymbol{x}-\boldsymbol{x}_{a} )
.\label{218}
\end{equation}
For any arbitrary distribution of \(n\) zeros \(\boldsymbol{x}_{a}\), the regular solution of \eqref{218} obeying the boundary condition for Higgs vacuum of
\(	\displaystyle
	\lim_{|\boldsymbol{x}|\to\infty}
	|\phi| =
	v_{0}^{2}
\)
has the magnetic flux
\begin{equation}
	\Phi_{B}
	=
	\int d^{2} \boldsymbol{x} \, B
	=
	- \frac{1}{g}
	\oint_{|\boldsymbol{x}|\to\infty} dx^{i} \,
	\epsilon^{ij} \partial_{j}
	\ln \frac{|\phi|}{
		\prod_{a=1}^{n}
		| \boldsymbol{x}
		- \boldsymbol{x}_{a}|
	}
	= \frac{2\pi n}{g}
.\label{225}
\end{equation}
Note that these values are the same as those of the homogeneous case \cite{Bogomolny:1975de}, independent of the detailed shapes of the inhomogeneity \(\sigma(\boldsymbol{x})\), since they are completely determined by the boundary conditions at both spatial infinity and \(n\) positions \(\boldsymbol{x}_{a}\).

In homogeneous case where \(v^{2}(\boldsymbol{x}) = v_{0}^{2}\), solving \eqref{218} is equivalent to solving two Bogomolny equations \eqref{209}--\eqref{210}.
Then, existence and uniqueness of topological vortex solutions of arbitrary shape with separated vorticities of the equation \eqref{218} are well-known \cite{Taubes:1979tm}.
For inhomogeneous case with \( \sigma(\boldsymbol{x}) \neq 0 \), however, we have only solutions with positive vorticities as mentioned above.
Thus, only half of the solutions to the same equation \eqref{218} are true BPS solutions in this case.

It is straightforward to check by calculating analytic index \cite{Weinberg:1979er} that every \(n\)-vortex solution of the Bogomolny equations \eqref{209}--\eqref{210}, if it exists, possesses \(2n\) zero modes identified with \(n\) positions in 2-dimensional plane.
Since no negative mode exists and all of the zero modes are identified by those of \(n\) separated topological BPS vortices of vorticity \(n\), this may support uniqueness of the BPS solution of vacuum of zero energy or non-interacting \(n\)-vortices despite of arbitrary inhomogeneity \eqref{204} as long as the boundary behavior of inhomogeneity \eqref{224} keeps the topology of a circle \( \mathrm{S}^1 \) at spatial infinity.
This mathematical question is addressed elsewhere in \cite{prep2}.

\subsection{Nonrelativistic version}

The analysis has been made for relativistic regime in the previous subsection but is indeed applicable to nonrelativistic systems as we shall see in this subsection.
Suppose that a collective order has the characteristic speed \(v_{\text{p}}\) much slower than the light speed \(c\) of electromagnetic waves, \( \dfrac{ v_{\text{p}} }{c} \ll 1 \), and the order parameter is depicted in terms of a complex scalar field \( \phi = \phi( t, \boldsymbol{x}, z ) \) of charge \(g\) and zero spin.
In the context of wave theory, it can describe a sound wave produced in a media and then propagating in another homogeneous media of different tension and mass density.
This kind of sound waves can typically obey the dispersion relation
\(
	\omega^{2} =
	v_{\text{p}}^{2} ( \boldsymbol{k}^{2}
	+ \lambda v_{0}^{2} )
\),
whose quantized excitation can become a species of phonon described by a scalar amplitude about the homogeneous broken vacuum \( \delta |\phi| = v_{0} - |\phi| \).%
\footnote{
	The dispersion relation says that the complex scalar field may describe a phonon but can not be a Cooper pair of two electrons in superconductor, whose dynamics is governed by the first order time derivative term
	\( \displaystyle
		i\hbar\bar{\phi}
		\frac{\partial}{\partial t}
		\phi
	\)
	\cite{Arovas}.
}

If an impurity is expressed by the inhomogeneity in \eqref{204}, the theory of such nonrelativistic complex scalar field coupled to electromagnetism possesses both the light speed \(c\) for electromagnetic waves and the nonrelativistic speed \( v_{\text{p}} \) for the scalar field. Dynamics is governed by the action written for spatial \(3\)-dimensional samples invariant under translations along the \(z\)-axis
\begin{equation}
	S = \int dt \int d^{2} \boldsymbol{x} \int dz \,
	\Big[
		- \frac{ \epsilon_{0} c^{2} }{4}
		F_{\mu\nu} F^{\mu\nu}
		- \overline{ \mathcal{D}_{\mu}\phi}\mathcal{D}^{\mu }\phi
		- \frac{\lambda}{4}
		( |\phi|^{2} - v^{2} (\boldsymbol{x}) )^{2}
		+ s \frac{g}{\hbar} \sigma (\boldsymbol{x}) B
	\Big]
,\label{227}
\end{equation}
where we use the SI unit system in this subsection and the gauge-covariant derivative is
\begin{equation}
	\mathcal{D}_{0} \phi
	=
	\Big(
		\frac{1}{v_\text{p}}
		\frac{\partial}{\partial t}
		+ i\frac{g}{\hbar c}
		\Phi
	\Big) \phi
,\qquad
	\mathcal{D}_{i} \phi
	= D_{i} \phi
	=
	\Big(
		\frac{\partial}{\partial x^{i}}
		- i \frac{g}{\hbar} A^{i}
	\Big) \phi
	.
\end{equation}
Note that the speed \(v_\text{p}\) enters in the time derivative.
The samples of consideration probably have the flat shape of a thin or thick slab and thus we separate planar variables \(\boldsymbol{x} = (x,y)\) from the thickness variable \(z\).

Let \(\mathcal{E}\) be the energy per unit length along the \(z\)-axis,
\begin{equation}
	\mathcal{E}
	= \frac{E}{\int dz}
	= \int d^{2} \boldsymbol{x} \,
	\Big[
	\frac{\epsilon_{0}}{2} 
	\big(
		\boldsymbol{E}^{2}
		+ c^{2} \boldsymbol{B}^{2}
	\big)
	+ |\mathcal{D}_{0} \phi|^{2}
	+ |\mathcal{D}_{i} \phi|^{2}
	+ \frac{\lambda}{4}
	( |\phi|^{2} - v^{2} (\boldsymbol{x}) )^{2}
	- \frac{g}{\hbar} \sigma(\boldsymbol{x}) B
	\Big]
.
\end{equation}
Then, in the Weyl gauge \(\Phi =0\), it coincides with the original Ginzburg-Landau free energy in the static homogeneous limit, which provides a macroscopic description of superconductivity \cite{Tinkham}
\begin{equation} 
	E =
	\int d^{2} \boldsymbol{x} \int dz \,
	\Big[
		\frac{\epsilon_{0} c^{2}}{2} 
		\boldsymbol{B}^{2}
		+ |\mathcal{D}_{i} \phi|^{2}
		+ \frac{\lambda}{4}
		( |\phi|^{2} - v_{0}^{2} )^{2}
	\Big]
.
\end{equation}

In the aforementioned BPS limit of the relativistic homogeneous abelian Higgs model with critical quartic coupling
\(	\displaystyle
	\lambda =
	\frac{2 g^2}
	{\epsilon_{0} c^{2} \hbar^{2}},
\)
the rest energies of gauge boson and Higgs boson are degenerate,
\(	\displaystyle
	m_{A_{\mu}} c^{2}
	=
	\sqrt{ \frac{2}{\epsilon_0} }
	| gv_{0} |
	=
	m_\text{H} c^{2}
\)
\eqref{203}.
In the present nonrelativistic case, this degeneracy corresponds to the equality of the London penetration depth \( \sqrt{2} \lambda_\text{L} \) and the correlation length \(\xi\)
\begin{equation}
	\sqrt{2} \lambda_\text{L}
	=
	\frac{ \sqrt{\epsilon_0} c \hbar }
	{ |gv_{0}| }
	=
	\xi
.\label{223}
\end{equation}

Since both the relativistic Lagrangian density \eqref{206} and its nonrelativistic counterpart \eqref{227} take the same mathematical form except for different propagation speeds \(v_\text{p}\) of complex scalar field and \(c\) of electromagnetic waves, we can reach the same BPS expression for static configurations in the nonrelativistic system,
\begin{equation}
	\mathcal{E} =
	\int d^{2} x \,
	\bigg[
		\big|
			(\mathcal{D}_{1} + is \mathcal{D}_{2}) \phi
		\big|^{2}
		+ \frac{\epsilon_{0} c^{2}}{2}
		\Big|
			B +
			s\frac{g}{\hbar}
			\frac{1}{\epsilon_{0} c^{2}}
			( |\phi|^{2} - v^{2} (\boldsymbol{x}) )
		\Big|^{2}
		+ s\frac{g}{\hbar} v_{0}^{2} B
	\bigg]
,
\end{equation}
where the sign can also be chosen as \(s=+1\) according to the same logic of the relativistic case.

The momentum density \(T\indices{^t_i} \) takes a similar form to the relativistic momentum density \eqref{228},
\begin{equation}
	T\indices{^t_i}
	=
	\frac{1}{c} T\indices{^0_i}
	=
	\epsilon_{0}
	( \boldsymbol{E} \times \boldsymbol{B} )^{i}
	+
	\frac{1}{v_\text{p}}
	(
		\overline{\mathcal{D}^{0} \phi}
		\mathcal{D}_{i} \phi
		+
		\overline{\mathcal{D}_{i} \phi}
		\mathcal{D}^{0} \phi
	)
	+
	\Big(
		\frac{c}{v_\text{p}} - 1
	\Big) A^{i} \rho
,
\end{equation}
where \( j^{0} = c\rho \).
Thus, any static objects do not carry angular momentum.
Also subsequent discussions on the energy-momentum tensor for the BPS configurations is unchanged in this nonrelativistic version.

As a consequence, we can conclude that the system can be classified into the type I
\(	\displaystyle
	\Big(
		\lambda <
		\frac{2g^{2}}
		{\hbar^{2} \epsilon_{0} c^{2}}
	\Big)
\)
and the type \Romtwo\
\(	\displaystyle
	\Big(
		\lambda >
		\frac{2g^{2}}
		{\hbar^{2} \epsilon_{0} c^{2}}
	\Big)
\)
superconductors depending on the length scales just as in the conventional nonlinear Schr{\"o}dinger matter \cite{Abrikosov:1956sx, Arovas, Tinkham}.
Furthermore, the length scales are unaffected by the presence of inhomogeneity as long as the constant piece is unchanged at the spatial asymptote.
It should be emphasized that the model of our consideration is a nonrelativistic model of a complex scalar field with a speed \(v_\text{p}\) of sound wave, yet accommodating the BPS structure by a straightforward application of the Bogomolny trick.

\section{Inhomogeneous Vacuum}

In usual field theories without inhomogeneity, the minimum energy solution is obtained by constant field configurations since any physical variation in spacetime costs energy and then the constant configuration is identified as the vacuum.
If the translation symmetry is broken, however, there is no a priori reason for that.
In this section, the inhomogeneous model \eqref{206} is taken into account and the vacuum configurations are explored.
The abelian Higgs model without inhomogeneity possesses obviously a symmetry-broken vacuum of constant scalar field \( \phi=v_{0} \).
Once inhomogeneity \( \sigma(\boldsymbol{x}) \) \eqref{204} is turned on for position-dependent \( v(\boldsymbol{x}) \), however, the field configuration of the broken vacuum cannot be left as a constant even in the BPS limit.
Furthermore, naive \(\phi = v(\boldsymbol{x}) \) configuration with vanishing electromagnetic field would not minimize the energy.

Recall that the energy is bounded from below by the magnetic flux as in \eqref{208} and thus any vacuum configuration should also satisfy Bogomolny equations \eqref{209}--\eqref{210} with vanishing magnetic flux \(\Phi_{B}=0\).
In order to find the would-be vacuum configuration, we need to examine \eqref{218} in the \(\Phi_{B}=0\) sector, i.e., without \(\delta\)-function terms in the right hand side,
\begin{equation}
	\nabla^{2} \ln |\phi|^{2}
	=
	2g^{2} [ |\phi|^{2} - v^{2} (\boldsymbol{x}) ]
.\label{301}
\end{equation}
Note that \(\phi = v(\boldsymbol{x})\) cannot be a solution for any nontrivial \(v(\boldsymbol{x})\) on \(\mathbb{R}^2\) with the boundary condition \( v(\boldsymbol{x}) \to v_{0} \) at spatial infinity.
Nevertheless, existence of a nontrivial vacuum solution satisfying \eqref{301} is physically evident at least for some reasonable inhomogeneity \(\sigma(\boldsymbol{x})\).
It can actually be proven that this is indeed the case, but mathematically rigorous proof is given elsewhere \cite{prep2}.

As a reasonable explicit example, we select \( \sigma(\boldsymbol{x}) \) of a Gaussian function with rotational symmetry
\begin{equation}
	\sigma (\boldsymbol{x})
	= - \beta v_{0}^{2} e^{ - \alpha^{2} m^{2} r^{2} }
,\label{302}
\end{equation}
where \( m = \sqrt{2}gv_{0} \) is the scalar mass \eqref{203} in \( |\boldsymbol{x}| \to\infty \) limit.
In nonrelativistic theory, \(m\) is replaced by
\(	\displaystyle
	\frac{mc}{\hbar}
	=
	\sqrt{\frac{2}{\epsilon_0}}
	\frac{gv_{0}}{\hbar c}.
\)
Then, this has a Gaussian dip (or bump) of its depth given by a dimensionless \(\beta\) at the origin and radial size controlled by another dimensionless parameter \(\alpha^{-1}\).
Note in particular that, for \(\beta=1\), the naive ``vacuum expectation value'' \( v(\boldsymbol{x}) \) of the broken vacuum vanishes at the origin.
To find the broken vacuum solution of \eqref{218} deformed from \(\phi = v_{0}\), we adopt a rotationally symmetric ansatz \(\phi = |\phi(r)|\) without any phase factor which makes \eqref{301} an ordinary differential equation
\begin{equation}
	\frac{ d^{2} \ln |\phi|^{2} }{d r^{2}}
	+ \frac{1}{r} \frac{d \ln |\phi|^{2} }{dr}
	=
	2g^{2} [
		|\phi|^{2} -
		v_{0}^{2} ( 1 - \beta e^{-\alpha^{2} m^{2} r^{2} } )
	]
.\label{303}
\end{equation}

The appropriate boundary conditions to be a broken vacuum are
\begin{align}
	\lim_{r\to\infty} \phi 
	& = v_{0}
, \\
	\lim_{r\to 0} \frac{d\phi}{dr}
	& = 0
.
\end{align}
Power series expansion near the origin consistent with \eqref{303} gives
\begin{equation}
\begin{aligned}[b]
	|\phi| \approx
	\phi_{0} v_{0}
	\bigg\{
		1
		& +
		\frac{1}{8}
		(\phi_{0}^{2} + \beta - 1)
		(mr)^{2}
\\
		& +
		\frac{1}{128}
		\bigg[
			(\phi_{0}^{2} + \beta - 1)^{2}
			+ \phi_{0}^{2} (\phi_{0}^{2} + \beta - 1)
			- 4 \alpha^{2} \beta
		\bigg]
		(mr)^{4}
		+ \cdots
	\bigg\}
,
\end{aligned}
\end{equation}
where \(\phi_{0}\) is a constant to be determined by proper asymptotic behavior.
At large distances, the asymptotic behaviors of the solution is independent of the inhomogeneous term,
\begin{equation}
	|\phi| \approx
	v_{0} [ 1 - \phi_{\infty} K_{0} (mr) ]
,
\end{equation}
where \(\phi_{\infty}\) is a constant to be determined by the behavior near the origin.

\begin{figure}[H]
	\centering
	\includegraphics[
		width=0.55\textwidth
	]{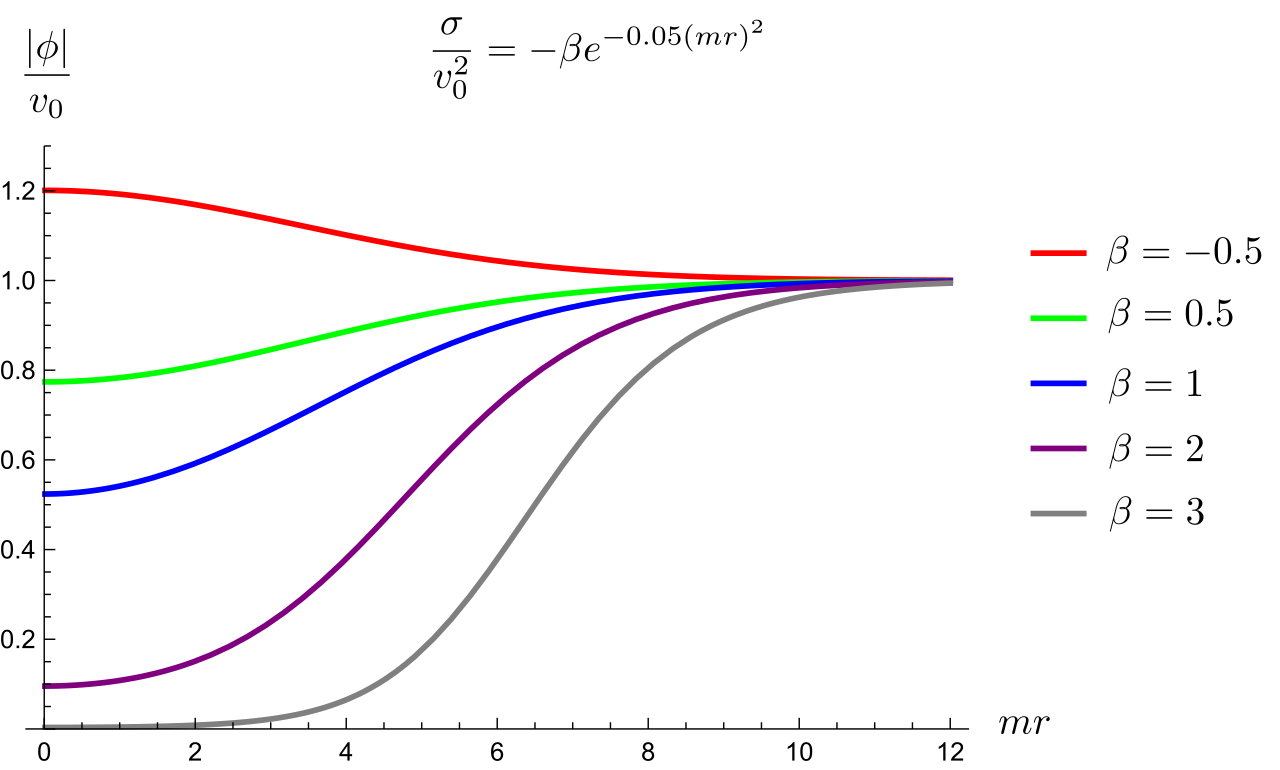}
	\caption{
		Scalar amplitudes \( |\phi| \) of the inhomogeneous broken vacuum for various depths \(\beta\) with fixed radial size \( \alpha^{-1} = 2\sqrt{5} \) in the unit of \(v_0\).
	}
\label{fig:301}
\end{figure}

Typical profiles of the inhomogeneous BPS vacuum configuration \(\phi\) obtained by numerical works are illustrated in Figure \ref{fig:301} for a fixed radial size \( \alpha^{-1} = 2\sqrt{5} \approx 4.5 \) and various depths \(\beta\).
As \(\beta\) increases, the Gaussian dip at the origin gets deeper and subsequently the profile of scalar field \(\phi\) in the neighborhood of the origin moves towards the \(\phi=0\) away from the asymptotic value (\(|\phi| = v_{0}\)).
Note that \( v^{2}(0) = 1 - \beta \) corresponds to zero for \(\beta=1\) and becomes negative for \(\beta>1\) which means \( |\phi|^{2} = v_{0}^{2} \) can no longer be the minimum of the potential around the origin.
There still exists a broken vacuum solution connecting smoothly two boundary conditions.

\begin{figure}[H]
	\centering
	\includegraphics[
		width=0.55\textwidth
	]{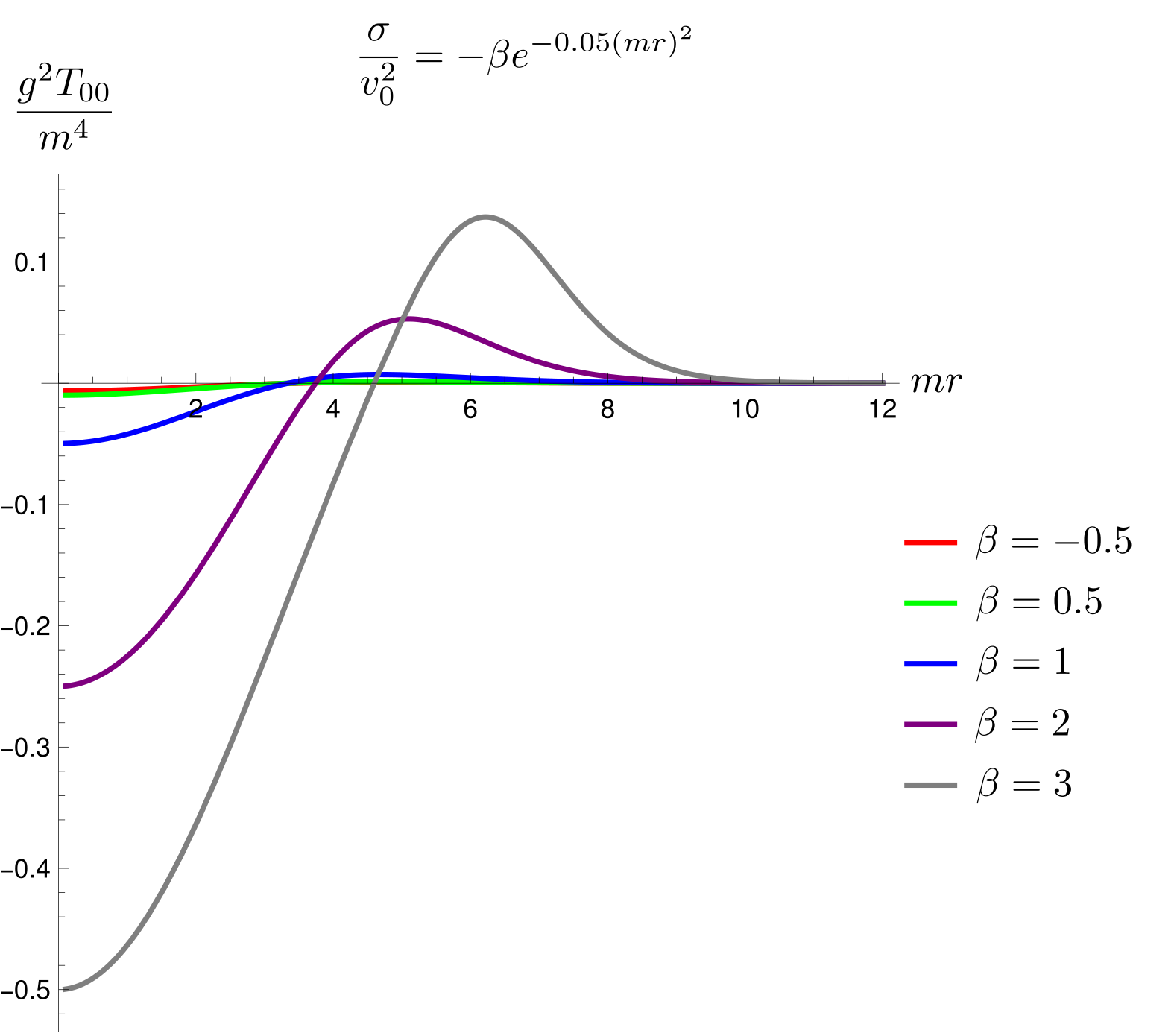}
	\caption{
		Energy densities \( T_{00} \) of the inhomogeneous broken vacuum for various depths \(\beta\) with fixed radial size \( \alpha^{-1} = 2\sqrt{5} \) in natural unit of \( m^{4}/ g^{2} \).
		In nonrelativistic theory, \(m^{4}/g^{2}\) is replaced by 
		\(	\displaystyle
			\frac{\epsilon_{0} c^{6}}{\hbar^2}
			\frac{m^4}{g^2}.
		\)
	}
\label{fig:304}
\end{figure}

The corresponding energy density \(T_{00} = - T\indices{^t_t}\) has negative minimum value of a bump around the origin and forms a ring-shaped region of positive energy density at a finite radius, whose depth and height deepen and become higher, respectively, as \(\beta\) increases (see Figure \ref{fig:304}).
It is worth recapitulating that the energy of this inhomogeneous BPS solution is indeed zero through a complete cancellation, as it should be.

\begin{figure}[H]
	\centering
	\includegraphics[
		width=0.55\textwidth
	]{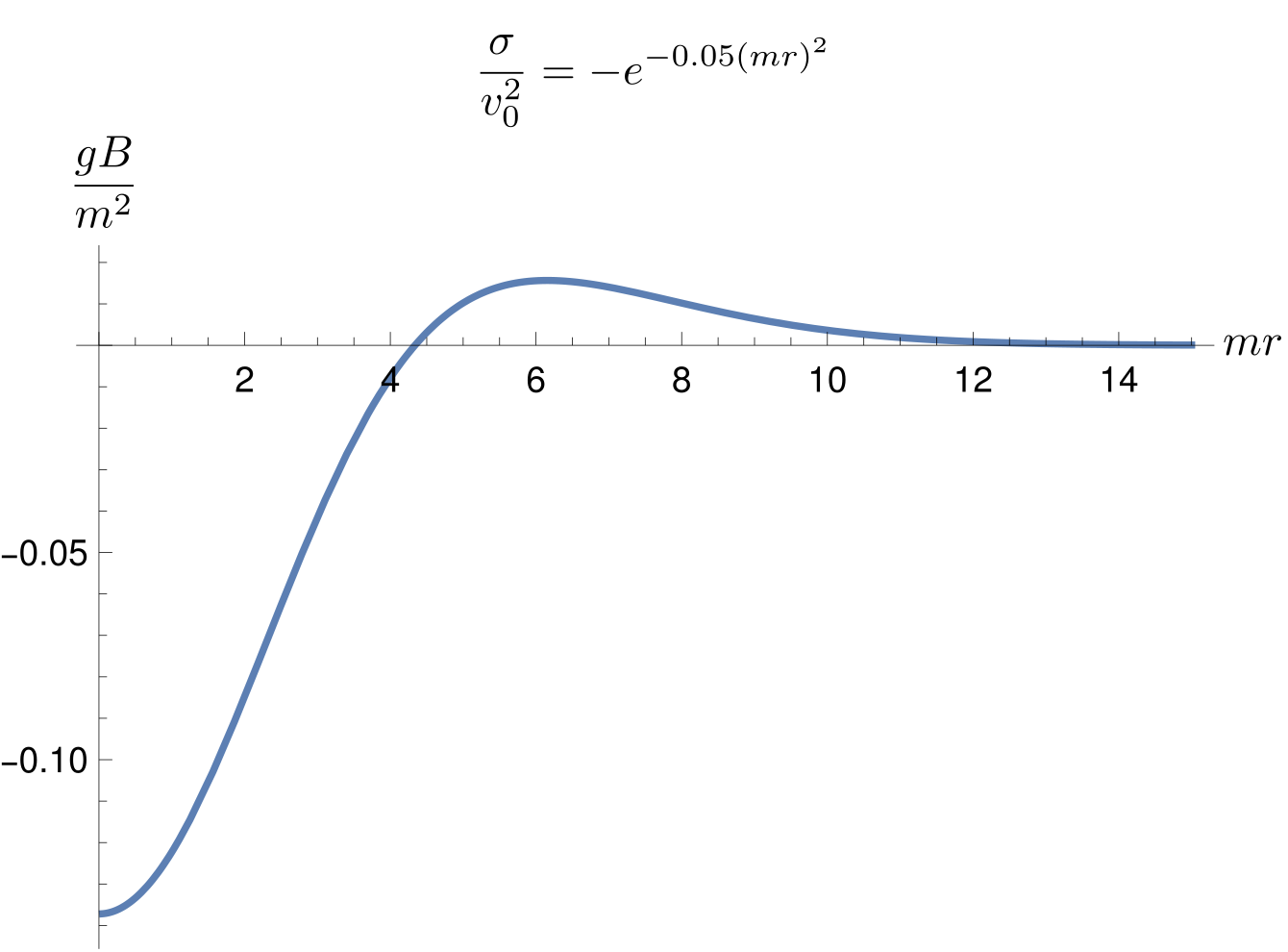}
	\caption{
		Magnetic field \(B\) of the inhomogeneous broken vacuum with unit depth \(\beta=1\) and radial size \( \alpha^{-1} = 2\sqrt{5} \) of the Gaussian type \( \sigma(r) \) in natural unit of \( m^{2}/g \).
		In nonrelativistic theory, \(m^{2}/g\) is replaced by 
		\(	\displaystyle
			\frac{c^2}{\hbar}
			\frac{m^2}{g}.
		\)
	}
\label{fig:302}
\end{figure}

Similarly, the magnetic field is nonzero since the vacuum solution is not given by \( |\phi|^{2} = v^{2} (\boldsymbol{x}) \).
As plotted in Figure \ref{fig:302}, the magnetic field increases from its negative minimum, turns its sign about \( mr \approx \alpha^{-1} \), and reaches a maximum of a positive ring-shaped profile.
The corresponding magnetic flux \(\Phi_{B}\) should be zero for the vacuum solution and we checked it numerically. 

\begin{figure}[H]
	\centering
	\subfigure[]{
		\includegraphics[
			width=0.48\textwidth
		]{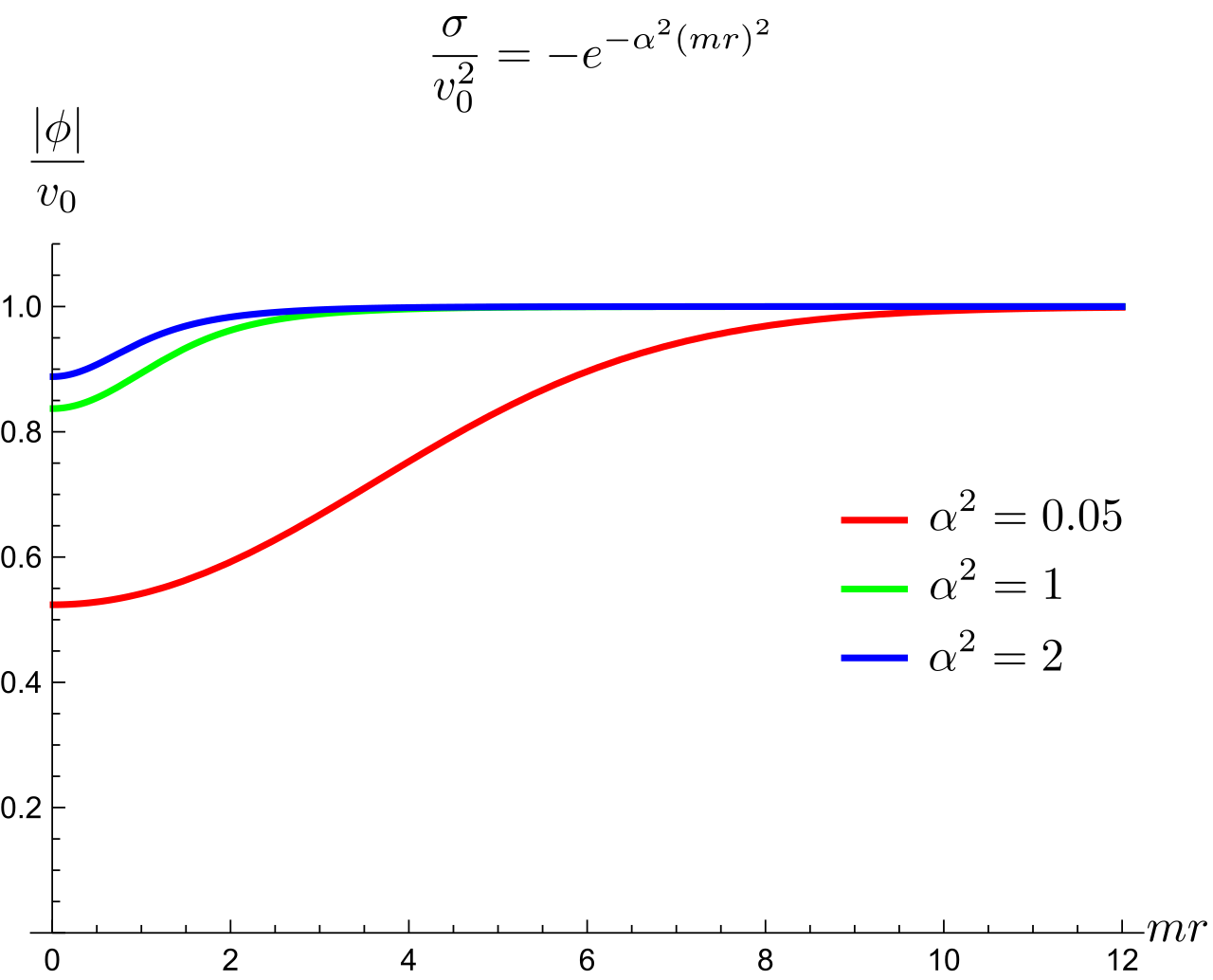}
	}
	\hfill
	\subfigure[]{
		\includegraphics[
			width=0.44\textwidth
		]{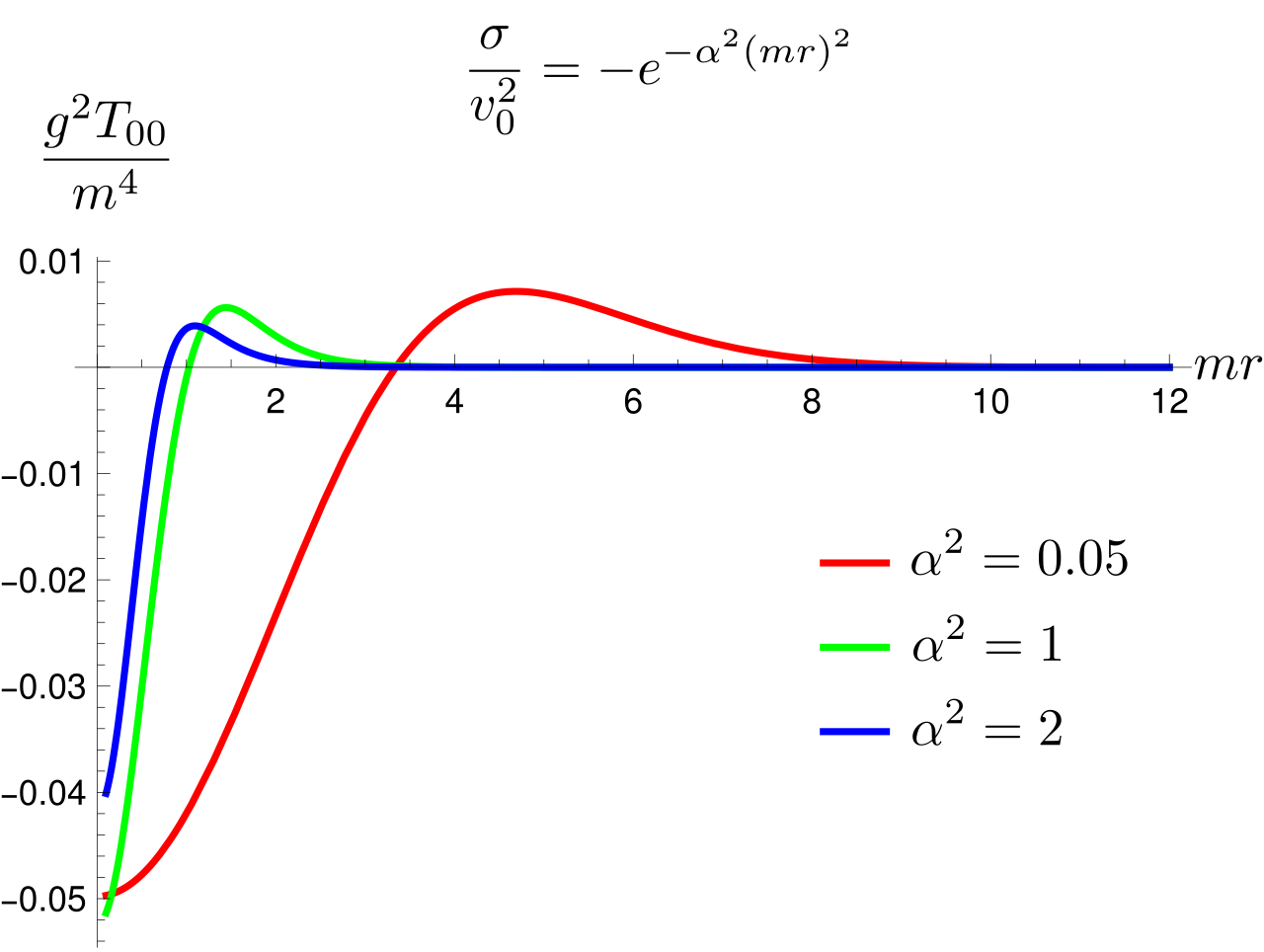}
	}
	\caption{
		(a) Scalar amplitudes \(|\phi|\)
		and
		(b) energy densities \( T_{00} \)
		of the inhomogeneous broken vacuum for various radial sizes \( \alpha^{-1} \) with fixed depth \(\beta=1\) in units of \(v_0\) and \( m^{4}/g^{2} \), respectively.
	}
\label{fig:303}
\end{figure}

Figure \ref{fig:301} shows that the radius of the region in which the scalar field \(\phi\) is substantially different from the asymptotic value \(v_0\) is determined not by the Higgs mass \eqref{203} of \( mr \approx 1 \) but by the size of inhomogeneous region, \( \alpha^{-1} \approx 4.5 \), since it is due to the inhomogeneous magnetic impurity term \eqref{205} added to the model.
As the size parameter \(\alpha^{-1}\) increases with a fixed \(\beta = 1\), the minimum value of \(\phi\) at the origin decreases and size of \(r\)-dependent region becomes larger as in Figure \ref{fig:303}.
The corresponding energy density \(T_{00}\) involves a negative energy bump around the origin and a positive energy ring-shaped region, of which the integrated energy always vanishes.

We conclude this section by considering the $\delta$-function limit of
the impurity~\eqref{302}. With $\beta = 4\alpha\eta$, $\sigma$ is rewritten
as
\begin{align} \label{delta}
\sigma(r) &= -4 \eta \alpha v_0^2 e^{-\alpha m^2 r^2} \nonumber \\
 &\equiv -\frac{2\pi\eta}{g^2} \delta_{\alpha}(r).
\end{align}
In the $\alpha \rightarrow \infty$ limit for fixed $\eta$,
$\delta_{\alpha}(r)$ becomes the two-dimensional delta function 
$\delta(\boldsymbol{x})$ and \eqref{301} leads to
\begin{equation}
\nabla^{2} \ln |\phi|^{2}
= 2g^{2} ( |\phi|^{2} - v_0^{2} ) + 4\pi\eta\delta(\boldsymbol{x}) .
\end{equation}
If $\eta$ is a positive integer, this equation is formally identical to 
the vortex equation \eqref{218} of the homogeneous theory with a
zero of multiplicity $\eta$ at the origin. Therefore,
the profile of the scalar amplitude $|\phi|$ is simply identical to that of
the vortex solution of the vorticity $\eta$ in the homogeneous theory discussed 
in the next section; see Figure~\ref{fig:401} for the explicit profile.
Note, however, that the inhomogeneous vacuum solution in the present case 
has no nontrivial phase unlike the vortex solutions produced under
the ansatz \eqref{401} in the homogeneous theory. 
If $\eta$ is positive but not an integer, the profile interpolates between 
those of the neighboring integer cases. Near the origin, the scalar field 
behaves as $\phi \sim r^\eta$ and thus vanishes at the position of 
the $\delta$-function impurity. Finally, if $\eta$ is negative, the scalar 
field diverges as $\phi \sim r^{-|\eta|}$ at $r=0$ and no regular vacuum 
solution is allowed. This is easily expected from the profile of $|\phi|$
in Figure~\ref{fig:301}, where $|\phi(0)|$ increases as $\beta$ becomes
smaller.

\section{Vortices with Inhomogeneous Mass}

In this section, we discuss positive energy solutions of the Bogomolny equations \eqref{209}--\eqref{210}.
In the homogeneous abelian Higgs model, topological vortices called the Abrikosov-Nielsen-Olesen vortices are found and are known to be noninteracting in the BPS limit of critical quartic coupling \(\lambda=1\) which is consistent with equal London penetration depth and correlation length \eqref{223} at least at tree level.
Since the inhomogeneity \( \sigma(\boldsymbol{x}) \) is assumed to be localized \eqref{224} and the \(\mathrm{U}(1)\) broken vacua are not altered, it is natural to expect that vortices are again supported in the inhomogeneous abelian Higgs model.
Moreover, with the critical quartic coupling, the BPS bound \eqref{208} implies that the vortices obtained by solving the Bogomolny equations \eqref{209}--\eqref{210} are left to be noninteracting.

In most of this section, we assume the rotationally symmetric Gaussian-type inhomogeneity \eqref{302} and look for rotationally symmetric solutions.
The ansatz for vortex configurations compatible with rotational symmetry is
\begin{equation}
	\phi =
	| \phi (r) | e^{in\theta}
, \quad
	A^{i} = 
	- \frac{1}{g} \epsilon^{ij} x_{j}
	\frac{ A_{\theta} (r) }{ r^2 }
,\label{401}
\end{equation}
where \(n\) stands for the vorticity of the solution and should be a positive integer since \(\phi\) is single-valued and we choose \(s=1\) as mentioned in section 2.1.
Then the Bogomolny equations \eqref{209}--\eqref{210} become
\begin{align}
	\frac{d |\phi|}{dr}
	& =
	\frac{1}{r} ( n - A_{\theta} ) |\phi|
,\label{402}
\\
	\frac{1}{r} \frac{d A_{\theta}  }{dr}
	& =
	g^{2} [ v_{0}^{2} + \sigma(r) - |\phi|^{2} ]
.\label{403}
\end{align}
Combination of these two equations by eliminating gauge field \(A_\theta\) leads to the rotationally symmetric version of the single second-order equation \eqref{218} for \(n\) vortices superimposed at the origin
\(
	\boldsymbol{x}_{a}
	= \boldsymbol{0}
	~ (a = 1,2,\cdots,n)
\).

For non-singular vortex solutions, the boundary conditions at the origin should be
\begin{equation}
	|\phi| (0) = 0
,\qquad
	A_{\theta}(0) = 0
.
\end{equation}
Finiteness of the energy requires the boundary conditions at spatial infinity
\begin{equation}
	|\phi|(\infty) = v_{0}
, \qquad
	A_{\theta} (\infty) = n
.       
\end{equation}
The boundary conditions of the gauge field confirm the quantized magnetic flux \eqref{225}
\begin{equation}
	\Phi_{B} 
	=
	2 \pi
	[ A_{\theta}(\infty) - A_{\theta}(0) ]
	= \frac{ 2\pi n}{g}
.\label{404}
\end{equation}
The energy of the solutions then saturates the Bogomolny bound \eqref{208}, i.e.,
\begin{equation}
	E
	= g v_{0}^{2} \Phi_{B}
	= 2 \pi v_{0}^{2} n,
\end{equation}
which suggests non-interacting nature of BPS vortices even in the presence of inhomogeneity.

Near the origin, the power series expansion of the solution gives
\begin{align}
	|\phi|
	& \approx
	\phi_{n} v_{0} (mr)^{n}
	\bigg[
		1
		- \frac{1 - \beta}{8} (mr)^{2}
		+ \frac{
			(1-\beta)^{2}
			-
			4 (
			\alpha^{2} \beta
			- \delta_{1n} \phi_{n}^{2}
			)
		}{128} (mr)^{4}
		+ \cdots
	\bigg]
, \\
	A_{\theta}
	& \approx
	\frac{1 - \beta}{4} (mr)^{2}
	+
	\frac{
		\alpha^{2} \beta
		- \delta_{1n} \phi_{n}^{2}
	}{8} (mr)^{4}
	+ \cdots
,
\end{align}
whose \(\phi_{n}\) is some constant whose specific values depend on \(n\).
At large distances, the asymptotic behaviors of the solutions are the same as those in homogeneous cases,
\begin{align}
	|\phi| & \approx
	v_{0} [ 1 - \phi_{n\infty} K_{0} (mr) ]
, \\
	A_{\theta} & \approx
	n + \phi_{n\infty} mr K_{1} (mr)
,
\end{align}
where \(\phi_{n\infty}\) is some constant.
Note that the behavior near the origin is completely different from the homogeneous case.
This is because the source term \eqref{302} with \(\beta=1\) cancels \(v_0\) at the origin so that \(v(0) = 0\), which invalidates the series expansion of the usual homogeneous case.
Also, if \(\beta > 1\), even the minimum of the potential at the origin changes from \( |\phi| = v_{0} \) to \( |\phi| = 0 \).
Despite the drastic change of the potential, topological vortex solutions can still be obtained by adjusting the parameters \(\phi_{n}\) or \(\phi_{n\infty}\).
In fact, if the parameter \(\phi_{n}\) is chosen too large, \(|\phi|\) reaches \(v_0\) at some finite \(r\) and then diverges.
If \(\phi_{n}\) is chosen too small, \(|\phi|\) goes to zero asymptotically.
Then there should exist a unique value of \(\phi_{n}\) for each vorticity \(n\) which satisfies the boundary condition \(|\phi(\infty)| = v_{0}\).

\begin{figure}[H]
	\centering
	\includegraphics[
		width=0.7\textwidth
	]{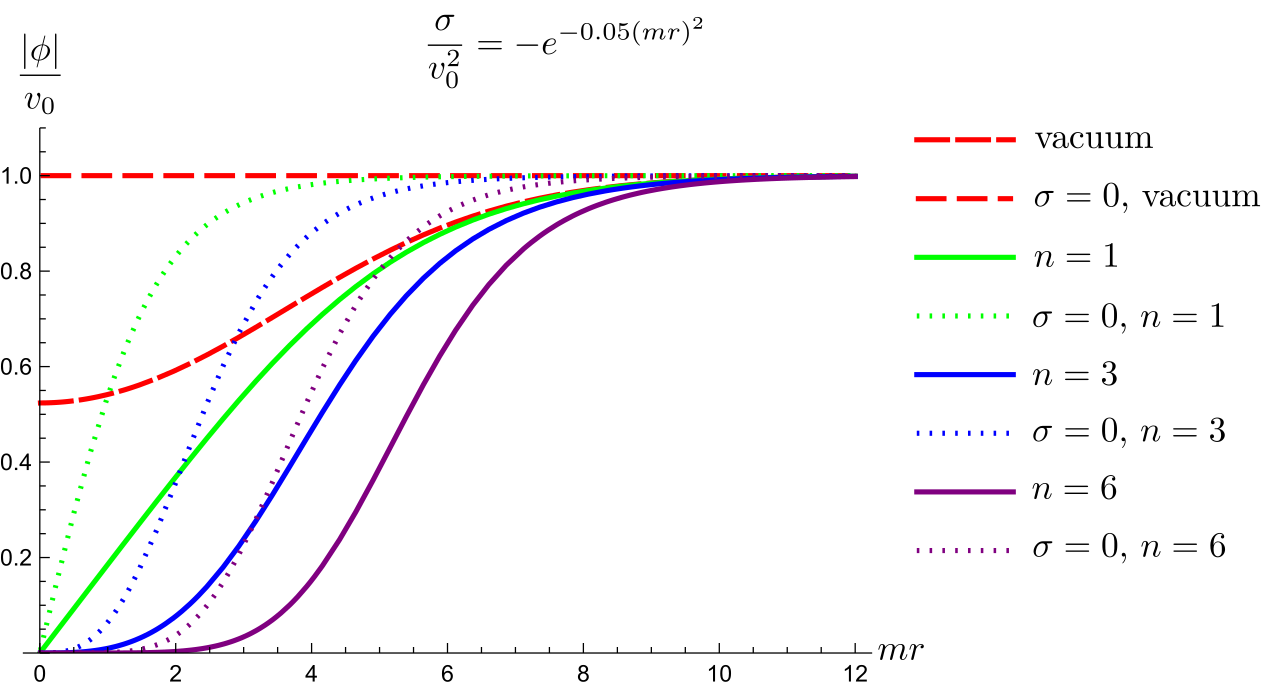}
	\caption{
		Scalar amplitudes \(|\phi|\) of topological vortices of vorticities \( n=1,3,6 \) (three solid curves) with fixed radial size \( \alpha^{-1} = 2\sqrt{5} \) and depth \( \beta=1 \) of the Gaussian type \( \sigma(r) \).
		The corresponding topological vortices in homogeneous limit (three dotted curves) and the homogeneous and inhomogeneous broken vacua (two dashed curves) are also plotted for comparison.
	}
\label{fig:401}
\end{figure}

Numerical solutions for \(n = 1, 3, 6\) with fixed \( \beta = 1 \) and \( \alpha^{-1} = 2\sqrt{5} \) are plotted in Figure \ref{fig:401} where the vacuum solution and solutions in the homogeneous case (\(\sigma=0\)) are also depicted for comparison.
We can easily see the effect of inhomogeneity \( \sigma(\boldsymbol{x}) \).

\begin{figure}[H]
	\centering
	\includegraphics[
		width=0.7\textwidth
	]{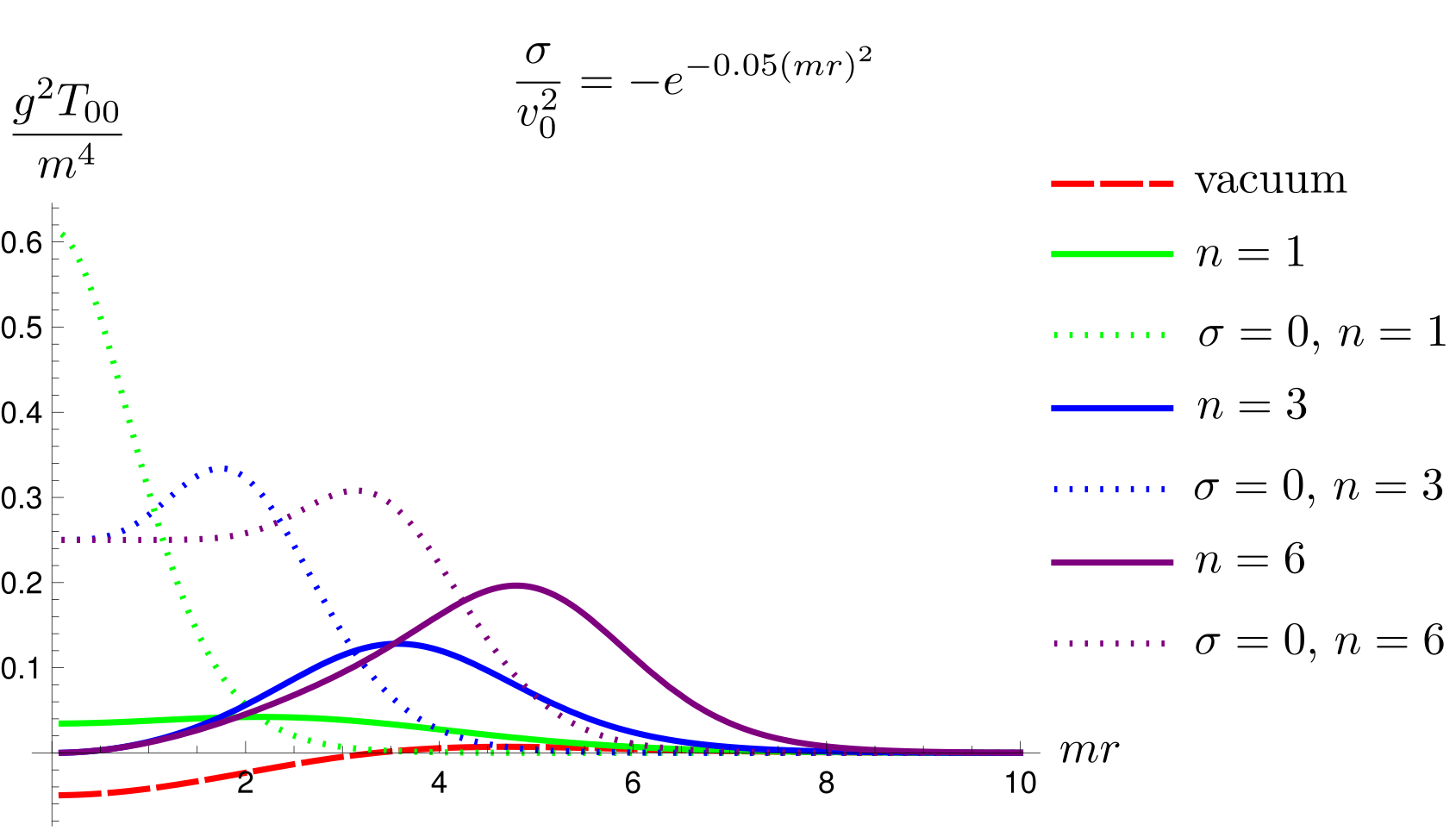}
	\caption{
	    Energy density \(T_{00}\) of rotationally symmetric vortex solutions of vorticities \( n=1,3,6 \) (solid curves) with fixed radial size \( \alpha^{-1} = 2\sqrt{5} \) and depth \( \beta=1 \) of the Gaussian type inhomogeneity \( \sigma(r) \) in natural unit of \( m^{4}/g^2 \).
		In nonrelativistic theory, \(m^{4}/g^{2}\) is replaced by 
		\(	\displaystyle
			\frac{\epsilon_{0} c^{6}}{\hbar^2}
			\frac{m^4}{g^2}.
		\)
	    The corresponding vortices in the homogeneous limit with \( \sigma = 0 \) (dotted curves) and the inhomogeneous broken vacuum (dashed curve) are also plotted for comparison.
	}
\label{fig:402}
\end{figure}

In Figure \ref{fig:402}, we plot energy density profiles of topological vortices with depth \(\beta = 1\) and radial size \( \alpha^{-1} = 2\sqrt{5} \). 
In this figure, we can clearly see the effect of \(\sigma(r)\) with \(\beta=1\).
Since \( \phi(0) = 0 \) for vortex solutions, the potential vanishes at the origin, i.e.
\(
	V(\phi(0))
	= g^{2} v^{2}(0)/2
	= 0
\)
which is in contrast to the homogeneous case where
\(
	V(\phi(0))
	= g^{2} v_{0}^{4}/2
\)
and hence \(T_{00}\) does not vanish at the origin.
Nevertheless, we confirmed within the numerical accuracy of order $10^{-5}$
that the energy of each vortex of vorticity \(n\) is given by \( E = 2\pi v_{0}^{2} n\).%
\footnote{In nonrelativistic theory, the energy of each vortex per unit length along the \(z\)-axis is given by \( \mathcal{E} = 2\pi v_{0}^{2} n \).
}

\begin{figure}[H]
	\begin{center}
	\includegraphics[
		width=0.7\textwidth,
	]{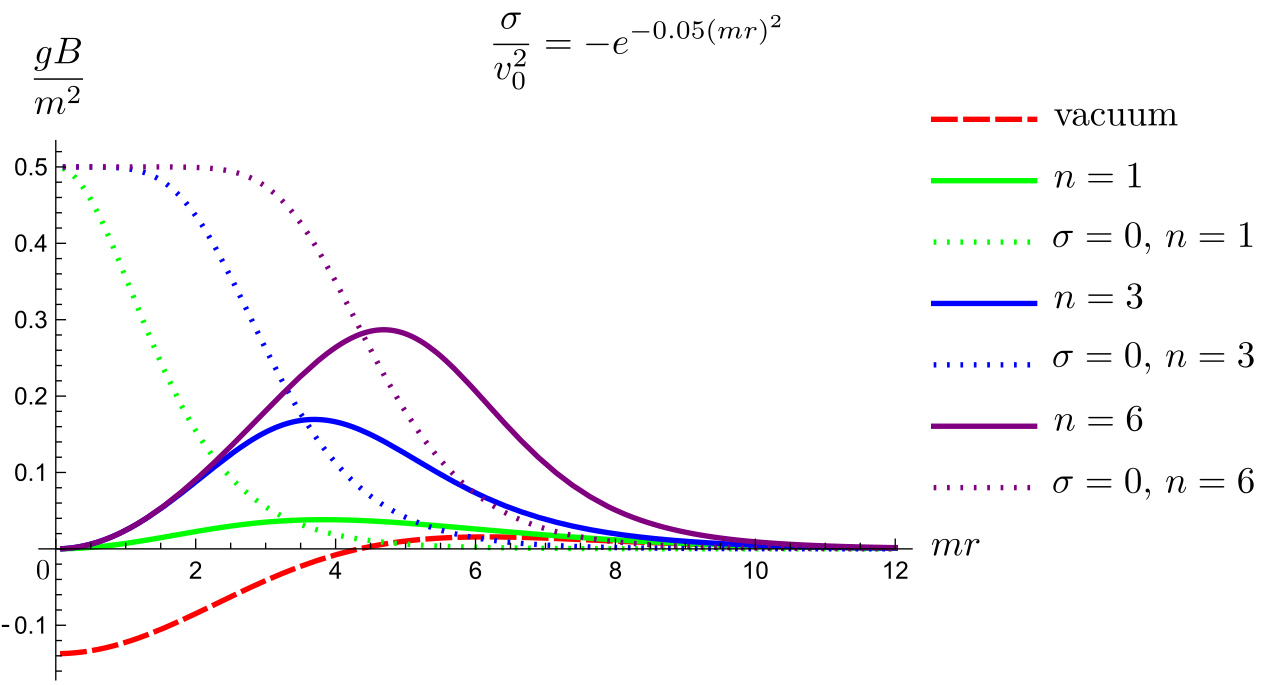}
	\end{center}
	\caption{
		Magnetic field \(B\) of rotationally symmetric vortices of vorticities \( n=1,3,6 \) (solid curves) with fixed radial size \( \alpha^{-1} = 2\sqrt{5} \) and depth \( \beta=1 \) of the Gaussian type \( \sigma(r) \) in natural unit of \( m^2/g\).
		In nonrelativistic theory, \(m^{2}/g\) is replaced by 
		\(	\displaystyle
			\frac{c^2}{\hbar}
			\frac{m^2}{g}.
		\)
		The corresponding topological vortices in homogeneous limit (dotted curves) and inhomogeneous broken vacuum (dashed curve) are also plotted for comparison.
	}
\label{fig:403}
\end{figure}

Figure \ref{fig:403} shows the magnetic fields of the vortex solutions which are ring-shaped in the presence of the inhomogeneity \( \sigma(r) \) with \( \beta = 1\).
This can easily be understood from the Bogomolny equation \eqref{210} since \( v(\boldsymbol{x}) \) vanishes at the origin with \( \beta = 1 \).
As in the case of energy, however, the magnetic fluxes have quantized values \( \Phi_{B} = 2\pi n /g \). 

Let us briefly comment on the vortex solutions in the $\delta$-function limit of
the impurity \eqref{delta}. As discussed towards the end of section 3, the
$\delta$-function impurity plays the role of an effective addition of
''vorticity'' $\eta$ at the impurity position as far as the scalar 
amplitude is concerned. For example, the rotationally
symmetric $n$-vortex solution behaves as $\phi \sim r^{n+\eta} e^{in\theta}$ 
near the origin, and hence the profile of $|\phi|$
mimics that of the homogeneous theory with vorticity $n+\eta$. 

Here, we focused on a specific Gaussian type of inhomogeneity \( \sigma(\boldsymbol{x}) \) given by \eqref{302}, although numerously potential types of impurities necessitate testing with various inhomogeneity profiles.
Ongoing work is exploring additional rotationally-symmetric impurities \(\sigma(r)\) \cite{prep1}.

\section{Conclusions and Discussions}

We have investigated the abelian Higgs model with a magnetic impurity in both relativistic and nonrelativistic regimes.
Our analysis is focused on the critical case where the quartic scalar coupling gives equal masses for gauge and Higgs bosons in the relativistic case, and the equality of the London penetration depth and correlation length in the nonrelativistic case.
We have examined the saturation of the BPS bound in detail.
While the usual homogeneous relativistic abelian Higgs model admits \( \mathcal{N}=2 \) supersymmetries, introduction of a magnetic impurity reduces this to \( \mathcal{N}=1 \).
We showed that symmetry-broken vacuum cannot be a trivial constant but has a nontrivial profile satisfying the Bogomolny equations, consistent with the breakdown of the translation symmetry.
Despite the spatial distribution of energy density and magnetic field, the total energy and magnetic flux remain zero.
In contrast to the homogeneous model, the inhomogeneous abelian Higgs model in the BPS limit is shown to allow only half of the solutions --- either nonnegative or nonpositive vorticities, but not both.
It should be noted that the nonrelativistic model based on a sound wave of slow propagation speed shares many characteristics with its relativistic counterpart, including the existence of BPS structure at the borderline of type I and \Romtwo\ superconductivity.
However, some distinct features emerge due to
the differences in time derivative term.

In order to gain insight into the properties of the new vacuum and vortices, we introduced a Gaussian impurity and investigated rotationally symmetric solutions of the Bogomolny equations.
In the region of inhomogeneity, local properties of the obtained BPS configurations are largely affected by the size and strength of the inhomogeneity.
However, in the spatial asymptote where homogeneity is nearly restored, the essential global characteristics of the vacuum and solitonic objects remain unaffected, thanks to topological protection.
We also considered the $\delta$-function limit of the Gaussian impurity and
showed that it plays the role of an effective addition of ``vorticity'' at the
impurity position in the homogeneous theory as far as the scalar amplitude
is concerned. It should also be noted that the impurity 
needs not be localized.
The study of long-ranged inhomogeneity could be intriguing as evidenced by the rich solitonic spectrum observed in \((1+1)\)-dimensional supersymmetric inhomogeneous theories \cite{Kwon:2021flc}.

This work considered only the abelian gauge field of electromagnetism whose dynamics is governed by the Maxwell term.
It would be beneficial to examine extended gauge symmetries \cite{Kim:1993mh} (e.g., \( \mathrm{U}(1)^{N} \) or nonabelian), or alternative kinetic terms such as the Chern-Simons term for planar physics \cite{ImCSH}, which is applicable to the fractional quantum Hall effect \cite{Tsui:1982yy}, or anyon superconductivity \cite{Chen:1989xs}.
The inclusion of other matter contents including fermions such as electrons should also be taken into account in the light of accompanying supersymmetry.
Moreover, meticulous efforts have been made to investigate the nonvanishing interactions in non-BPS structures away from the critical coupling, e.g. \( \lambda \neq 2 g^{2} \) in the natural unit system for relativistic abelian Higgs model and
\(	\displaystyle
	\lambda \neq
	\frac{2g^2}
	{\epsilon_{0} c^{2} \hbar^{2}}
\)
in the SI unit system for the nonrelativistic abelian Higgs model.
We anticipate that our findings will contribute to field-theoretic description of realistic dirty planar and bulk samples and that they will facilitate precise, systematic studies on various kinds of impurities in the framework of field theories.

\section*{Acknowledgement}

This work was supported by the National Research Foundation of Korea(NRF) grant with grant number NRF-2022R1F1A1074051 (C.K.), NRF-2022R1F1A1073053 (Y.K.), and NRF-2019R1A6A1A10073079, RS-2023-00249608 (O.K.).

\end{document}